 \documentclass[final,1p,times,]{elsarticle}
\journal{Physica A}
\usepackage[all]{nowidow}
\usepackage{csquotes}

\biboptions{sort&compress}

\usepackage{mathtools}

\usepackage[detect-all]{siunitx}
\usepackage{bm}
\usepackage{amsmath}
\usepackage{bm}
\usepackage{xspace}

\usepackage     [utf8]                  {inputenc}
\usepackage     [T1]                    {fontenc}
\usepackage                             {color}
\usepackage                             {amsmath}
\usepackage                             {physics}
\usepackage[margin=10pt,font=small,labelfont=bf,format=plain]{caption}
\usepackage{siunitx}
	\DeclareSIUnit{\c}{\text{\ensuremath{c}}}
	\DeclareSIUnit{\fm}{\femto\meter}
\usepackage                             {amssymb}
\usepackage[thinc]{esdiff}
\usepackage{mathtools}
\usepackage [version=4]                 {mhchem}
\usepackage                             {graphicx}
\usepackage                             {subfig}
\usepackage                             {scrlayer-scrpage}

\usepackage     [english]               {babel}
\usepackage                             {svg}
\usepackage{csquotes}
\usepackage{lmodern}
\usepackage                             {hyperref}
\usepackage                             {cleveref}

\definecolor{darkblue}{rgb}{0.0,0.0,0.4}
\definecolor{darkgreen}{rgb}{0.0,0.4,0.0}
\hypersetup{
    colorlinks,
    linkcolor=black,
    citecolor=darkgreen,
    urlcolor=darkblue
}

\usepackage     [utf8]                  {inputenc}
\usepackage     [T1]                    {fontenc}
\usepackage                             {color}
\usepackage                             {amsmath}
\usepackage                             {physics}
\usepackage[margin=10pt,font=small,labelfont=bf,format=plain]{caption}
\usepackage{siunitx}
	\DeclareSIUnit{\c}{\text{\ensuremath{c}}}
	\DeclareSIUnit{\fm}{\femto\meter}
\usepackage                             {amssymb}
\usepackage[thinc]{esdiff}
\usepackage{mathtools}
\usepackage [version=4]                 {mhchem}
\usepackage                             {graphicx}
\usepackage                             {subfig}
\usepackage                             {scrlayer-scrpage}

\usepackage     [english]               {babel}
\usepackage                             {svg}
\usepackage{csquotes}
\usepackage{lmodern}
\usepackage                             {hyperref}
\usepackage                             {cleveref}

\definecolor{darkblue}{rgb}{0.0,0.0,0.4}
\definecolor{darkgreen}{rgb}{0.0,0.4,0.0}
\hypersetup{
    colorlinks,
    linkcolor=black,
    citecolor=darkgreen,
    urlcolor=darkblue
}

\begin{document}

\begin{frontmatter}



\title{Numerical solution of the nonlinear boson diffusion equation for gluons}


\author{J.\,R\"ossler and G.\,Wolschin\corref{cor}}
\ead{g.wolschin@thphys.uni-heidelberg.de}
\address{Institut f{\"ur} Theoretische Physik der Universit{\"a}t Heidelberg, Philosophenweg 16, D-69120 Heidelberg, Germany, European Union}

%\cortext[cor]{Corresponding author}

\begin{abstract}
The nonlinear boson diffusion equation is taken as a basis to account for the fast thermalization of gluons in the initial stages of relativistic heavy-ion collisions. For constant drift and diffusion coefficients with schematic initial conditions, this equation has previously been solved exactly. In order to achieve a more realistic time evolution towards thermalization, energy-dependent transport coefficients are introduced, requiring numerical solutions of the nonlinear equation. Their accuracy is tested against the exact analytical results in the limit of constant coefficients.
%The physical motivation for energy-dependent transport coefficients is discussed, the numerical solution is developed, differences between solutions for constant and energy-%dependent transport coefficients are investigated, and the 
The consequences for transient gluon-condensate formation through elastic scatterings in overoccupied systems are discussed.\\
\end{abstract}
 

\end{frontmatter}
%\newpage

\section{Introduction}
\label{intro}

In relativistic heavy-ion collisions of spatially extended nuclei such as gold or lead, the initial parton distributions in momentum space at $t=0$ are approximately described by $\theta$-functions: for valence quarks up to the Fermi momentum, for gluons up to the gluon saturation scale $Q_\text{s}\simeq1$\,GeV \cite{mue00}. During the short initial nonequilibrium stages of the collision that last about 1 fm, both distribution functions quickly approach local statistical equilibrium: quarks reach the Fermi--Dirac limit, gluons the Bose--Einstein thermal distribution. The latter thermalize faster than quarks in about 0.1 fm, because they need not obey Pauli's principle, and have a larger color factor as compared to quarks.

Whereas the short-time local thermalization is a sufficient prerequisite for viscous hydrodynamic models or diffusion models of the main collision phase up to hadronization at $t\simeq 8-10$ fm, it does not imply that macroscopic variables like the net-proton (proton minus antiproton) distribution that is indicative of the stopping process
reach equilibrium: these distributions remain highly nonequilibrium, whereas transverse-momentum distributions of produced charged hadrons are observed to be very close to equilibrium.

There has been an abundant body of anterior works that have studied the long-standing issue of local thermalization of gluons, which has sometimes been considered as the single most important problem in the physics of relativistic heavy-ion collisions \cite{ba01}. Effective kinetic theories have been formulated that are, in particular, based on the quantum Boltzmann equation \cite{km11,jpb12,jpb13,kur14,blmt17,fuku17}, and many other papers have been quoted in these works. It has been a particularly intriguing idea that in overoccupied systems, particle-number conserving elastic collisions may trigger the transient formation of a gluon condensate \cite{jpb12,jpb13}. Accordingly, quantum Boltzmann equations for elastic scattering -- with a focus on overoccupied systems with transient condensate formation  -- have been solved \cite{scar14,mei16,gr17}. However, it turned out that the shorter timescale of inelastic collisions effectively prevents condensate formation \cite{blmt17,gr19}.

As a complement to these extensive numerical calculations, the fast thermalization of partons in the initial stages of relativistic heavy-ion collisions towards local statistical equilibrium has been modelled schematically based on a nonlinear diffusion equation for the occupation-number distributions in the full momentum range \cite{gw18,gw22}, as an approximation to the quantum Boltzmann equation.
Whereas the latter as well as other approximate kinetic equations \cite{jpb12,blmt17} that are using the small-angle approximation for gluon scattering \cite{jpb13} must be solved numerically, the nonlinear boson diffusion equation (NBDE) can be solved analytically in the limit of constant transport coefficients. It thus provides a transparent and analytically tractable model to account for aspects of the kinetic evolution of a dense parton system during the early moments of a relativistic heavy-ion collision. It is one of the few nonlinear partial differential equations with a clear physical meaning that can be solved exactly. Already in the simplified case of constant coefficients, results for the time-dependent occupation-number distribution turned out to be very similar to the numerical solutions of a more complicated Boltzmann-type equation in the small-angle scattering approximation \cite{blmt17} -- which is, however, only valid in the infrared region, whereas NBDE solutions exist for the full momentum range. 
 
%Analytical solutions of the nonlinear equation have been derived for both, fermions (quarks) \cite{gw82,gw18} and bosons (gluons) \cite{gw22}.  Although these solutions still depend %on the specific form of the initial conditions, it has been possible to derive exact analytical results for quarks \cite{bgw19} and gluons \cite{gw22a,mgw24} in the limit of constant %transport coefficients through a nonlinear transformation. For gluons, these are more difficult to obtain, because the boundary condition at the singularity that occurs at the chemical %potential $\mu$ must be considered in a combined initial- and boundary-value problem. It has been solved analytically in \cite{gw22a} for inelastic gluon scattering with $\mu=0$ and %$\theta$-function initial conditions. 

An exact, fully analytical solution of the NBDE has been performed in \cite{mgw24}, where we focused on particle-number conserving elastic scattering and the ensuing transient time-dependent gluon condensate formation in overoccupied systems. This problem remains physically interesting, in spite of the fact that inelastic collisions occur faster and actually prevent condensate formation to occur. When calculating the time-dependent condensate fraction in the limit of constant transport coefficients, however, it turns out that an overshoot of the equilibrium limit occurs as an artefact of the constant-coefficient approximation, because thermalization in the infrared (IR) is too fast compared to the one in the ultraviolet (UV) region, such that too many particles are pushed into the condensed state, thus overshooting the thermal limit. Hence, energy-dependent coefficients are required to prevent an overshoot, and also for basic physical reasons because the diffusion coefficient should vanish at zero energy, and also in the limit of large energies.

%It is the aim of our present work to extend the nonlinear diffusion approach to energy-dependent transport coefficients, which are %physically more realistic than constant ones. In particular, one expects drift and diffusion coefficients to vanish both at zero energy %(no diffusion in the condensed state), and in the ultraviolet (UV). As an additional benefit, an artifact in the time-dependent %condensate fraction that occurs in case of constant coefficients can be remedied when using energy-dependent ones: As we found %in \cite{mgw24}, for constant coefficients the condensate fraction slightly overshoots the equilibrium limit due to an unbalanced %equilibration in the infrared (IR) and UV regions. 

For energy-dependent transport coefficients, it is presently impossible to solve the nonlinear boson diffusion equation analytically. Hence, we resort to numerical solutions using the \textsc{Julia}  programming language. Whereas this may not seem innovative as compared to the many available numerical models mentioned above, it has the distinct advantage that in the limiting case of constant coefficients, it is possible to compare the numerical solutions to the exact results in order to assess how accurate the numerical approach turns out to be, before proceeding to energy-dependent coefficients. 

We focus on the thermalization of massless gluons with single-particle energy $\epsilon=|{p}|=p$ through elastic scatterings in overoccupied systems. This has been widely discussed in the literature starting with \cite{jpb12,jpb13}, where the possible occurence of gluon-condensate formation in relativistic heavy-ion collisions has first been proposed. 
%It has, however, meanwhile turned out that transient  condensate formation in relativistic nuclear collisions is unlikely, because the competing thermalization process of inelastic gluon scatterings (splitting and fusion) is significantly faster \cite{blmt17}. Nevertheless,
It is of principle interest to not only investigate the thermalization process, but also to explore whether our schematic model yields physically reasonable results for time-dependent gluon condensate formation through particle-number conserving elastic scatterings -- even though inelastic scatterings effectively prevent that to happen.

The nonlinear momentum-space diffusion model for massless gluons is briefly reviewed in the next section, including the time-dependent solutions of the NBDE  for constant transport coefficients $v, D$ that we had derived earlier. In Section\,\ref{num}, the numerical solution of the nonlinear diffusion equation is presented, which can be applied to both, the NBDE for constant coefficients -- including comparisons with the exact results --, and in Section\,\ref{energy_dep} using energy-dependent drift and diffusion coefficients. Results for the distribution functions of the gluonic single-particle occupation-number probability distributions are compared for constant and energy-dependent coefficients, and we discuss the effect of their energy dependence on the single-particle gluon distribution functions.  In Section\,\ref{cond}, we calculate the time-dependent condensate fraction for gluons in a central Pb-Pb collision close to a typical LHC center-of-mass energy per particle pair of $\sqrt{s_\text{NN}}=2.76$ TeV for both, constant and energy-dependent coefficients. The conclusions are drawn in Section\,\ref{conc}.

\section{Nonlinear diffusion model for massless gluons}
\label{model}
The basic nonlinear diffusion equation
 for the time-dependent single-particle occupation-number probability distributions $n\equiv n\,(\epsilon,t)$ of partons had been derived
 from the quantum Boltzmann collision term in \cite{gw18,gw22} as
 \begin{equation}
	%\frac{\partial n}{\partial t}=-\frac{\partial}{\partial{\epsilon}}\left[v\, n\,(1\pm n)+n\frac{\partial D}{\partial \epsilon}\right]+\frac{\partial^2}{\partial{\epsilon}^2}\bigl [D\,n\,\bigr]
	{\partial_t n}=-{\partial_{\epsilon}}\left[v\, n\,(1\pm n)+n\,{\partial_\epsilon D}\right]+{\partial_{\epsilon\epsilon}}\bigl [D\,n\,\bigr]
\label{nbde}
\end{equation}
where the $+$ sign represents bosons, and the $-$ sign fermions. The drift term $v\,(\epsilon,t)<0$ produces a shift of the distributions towards the infrared, and the diffusion function $D\,(\epsilon,t)$ causes a broadening with increasing time. The detailed properties of the many-body system are hidden in the drift and diffusion functions \cite{gw18,gw22}, which differ for elastic and inelastic processes. The first derivative-term of the diffusion coefficient ensures \cite{gw22}
% as compared to ref.\,\cite{gw18} 
that the stationary solutions $n^\pm_\infty(\epsilon)$ become Bose--Einstein or Fermi--Dirac equilibrium distributions, respectively, not only in case of constant coefficients $v, D$, but also for energy-dependent coefficients. The equilibrium distributions are then attained for $t\rightarrow\infty$ as
\begin{equation}
n_\infty^\pm(\epsilon)=n_\mathrm{eq}^\pm(\epsilon)=\frac{1}{e^{(\epsilon-\mu)/T}\mp 1}\,,
 \label{Bose--Einstein}
\end{equation}
provided the ratio of drift to diffusion has no energy dependence, and requiring that
%for the limit of time to infinity
$\lim_{t\rightarrow \infty}[-v\,(\epsilon,t)/D\,(\epsilon,t)] \equiv 1/T$ with the equilibrium temperature $T$. (We use units $k_\text{B}=\hbar=c=1$ throughout this manuscript).
The chemical potential is $\mu\leq0$ in a finite Bose system, and $\mu=\epsilon_\mathrm{f}>0$ for a Fermi system. It appears as an integration constant in the stationary solution of 
Eq.\,(\ref{nbde}) for $t\rightarrow\infty$. It vanishes in case of inelastic collisions that do not conserve particle number such that the temperature alone determines the equilibrium distribution. 

As in our previous article \cite{mgw24}, we concentrate on the nonlinear diffusion equation for bosons, to account for gluon thermalization. We focus on particle-number conserving elastic collisions, which may cause gluon-condensate formation in an overoccupied system \cite{jpb12} where the initial gluon content is larger than that of the final equilibrium distribution.

%with transport coefficients differing in both cases. Whereas these should be calculated from microscopic theory, we shall estimate them in this work on macroscopic grounds as in %\cite{gw22}, because our emphasis is on the exact analytical solution of Eq.\,(\ref{n}).
We use the ergodic approximation, where the occupation-number distributions 
$n\,(\epsilon,t)$ depend only on energy $\epsilon$ and time $t$, with  
 the energy dispersion relation in relativistic systems given by
%footnote{~$\hbar=c=1$; time units are written as fm, momenta as GeV} 
$\epsilon=\sqrt{|{p}|^2+m^2}$ for particles with mass $m$. We use in the following
$\epsilon=|{p}|=p$ for massless gluons.

For constant transport coefficients, the NBDE\,(\ref{nbde}) simplifies to
 \begin{equation}
	{\partial_t n}\,(\epsilon,t)=-v\,{\partial_{\epsilon}}\left[\, n\,(\epsilon,t)(1+n\,(\epsilon,t))\right]+D\,{\partial_{\epsilon\epsilon}}\bigl [n\,(\epsilon,t)\bigr]\,.	
	\label{bose}
\end{equation}
As shown in previous works \cite{gw18,gw22}, it can be solved analytically,
with the equilibrium temperature $T=-D/v$, and the chemical potential $\mu\le0$ appearing as an integration constant.
The simplified NBDE still preserves the essential features of quantum statistics that are contained in the quantum Boltzmann equation, such as Bose enhancement in the low-momentum region that increases rapidly with time. Its solutions have been applied at very low energies to elastic scattering of ultracold bosonic atoms, properly accounting for thermalization and time-dependent Bose--Einstein condensate formation \cite{gw22,kgw22} in agreement with deep-quench data for bosonic atoms. 

The diffusion equation with constant coefficients can be solved in closed form using the nonlinear transformation outlined in \cite{gw18,gw22}
\begin{equation}
	n\,(\epsilon,t) =T {\partial_{\epsilon}}\ln{{Z}(\epsilon,t)} -\frac{1}{2}= \frac{T}{{Z}} {\partial_\epsilon{Z}} -\frac{1}{2}
	\label{eq:Nformula}	
\end{equation} for any given initial condition $n_\mathrm{i}\,(\epsilon)$ with the time-dependent partition function $Z(\epsilon,t)$ that fulfills the linear diffusion equation
  \begin{equation}
\partial_t\,{Z}(\epsilon,t)=D\, {\partial_{\epsilon\epsilon}}{\,{Z}(\epsilon,t)}\,.
    \label{eq:diffusionequation}
\end{equation}
For bosons, one must consider the boundary conditions occuring at the singularity $\epsilon=\mu\le 0$ \cite{gw22}. No corresponding singularity exists for fermions, such that the fermionic exact solution of the nonlinear problem can be obtained with the free Green's function \cite{gw18,bgw19}. The bosonic partition function with boundary conditions becomes 
\cite{gw22}
      \begin{equation}
Z(\epsilon,t)= \int_0^{+\infty}{G}\,(\epsilon,x,t)\,F\,(x+\mu)\,\text{d}x\,,
    \label{eq:partitionfunctionZb}
    \end{equation}
    with the bounded Green's function $G(\epsilon,x,t)$ of the linear diffusion equation, and $F(x)$ an integral over the initial distribution $n_\mathrm{i}(x) = n_\mathrm{i}^0\theta(1-x/Q_\mathrm{s})$\\
    \begin{gather}
    G(\epsilon, x, t) =\frac{1}{\sqrt{4\pi Dt}}\left[ \exp\left(-\frac{(\epsilon-\mu-x)^2}{4Dt}\right)
    -\exp\left(-\frac{(\epsilon-\mu+x)^2}{4Dt}\right)\right] \,, \\
    F(x) = \exp\left[-\frac{1}{2D}\left(vx+2v\int^x n_\mathrm{i}(y)\,\mathrm{d}y\right)\right].
\end{gather}
The integration constant in the indefinite $y$-integral can be set to zero since it cancels out when taking the logarithmic derivative, and the time-dependent prefactor in Green's function can be omitted for the same reason. The occupation-number distribution that is calculated from the nonlinear transformation Eq.\,(\ref{eq:Nformula}) becomes infinity as the energy approaches $\mu$,  \(\lim_{\epsilon \downarrow \mu} n\,(\epsilon,t) = \infty\) \,$\forall$ \(t\). It attains the Bose--Einstein limit over the full energy range as $t\rightarrow \infty$ \cite{gw22}, not only in the thermal tail as in case of the free solutions \cite{gw18}.
The energy range for the bounded solution is restricted to $\epsilon \ge \mu$, and the integral in Eq.\, (\ref{eq:partitionfunctionZb}) can either be calculated numerically \cite{gw22,gw22a}, or analytically, with explicit results for $Z(\epsilon,t)$ and $\partial Z(\epsilon,t)/\partial\epsilon$ given in \cite{mgw24}.  {Since the nonlinear diffusion equation has the correct Bose--Einstein limit, it is expected that its applicability extends over the 
full physically relevant time range}. We shall later compare the ensuing analytical results for the time-dependent occupation-number distribution functions $n\,(\epsilon,t)$
with those from a fully numerical solution of the nonlinear boson diffusion equation for constant transport coefficients, and then
proceed to the case of energy-dependent coefficients.
\section{Numerical solution of the NBDE}
\label{num}
%\subsection{Solution for constant coefficients}
We use the method of lines \cite{MOLarticle} to numerically solve the partial differential equation. This involves a semi-discretization: all but one dimension (in our case, the time $t$) will be discretized using a symmetric finite-differences approach such that the result is a system of ordinary differential equations for the time $t$. For a uniform grid $x = \{x_i\,|\,i\in 1,\dots, \hat{N}\}$, with $\hat{N}$ cells and spacing $\Delta x$, the first- and second-order derivative of a differentiable function $f(x_i)\equiv f_i$ thus becomes
% Starting at the Taylor expansion of a function $f(x)$ around $f(x\pm \Delta x)$ for small $\Delta x$
% \begin{equation}\label{eq:Taylor}
%    f(x\pm\Delta x) \approx f(x)\pm \dv{f(x)}{x}\Delta x+\dv{f(x)}{x^2}\Delta x^2\pm\mathcal{O}(\Delta x^3)\,,
%     %  f(x\pm\Delta x) \approx f(x)\pm \dv{f(x)}{x}\Delta x+\dv{f(x)}{x^2}\Delta x^2\pm{O}(\Delta x^3).
% \end{equation}
% and subtracting the two equations \eqref{eq:Taylor}, one gets the discrete and symmetric first order derivative of $f(x)$
% \begin{gather}
%     f(x+\Delta x)-f(x-\Delta x) = 2\Delta x\dv{f(x)}{x}+\mathcal{O}(\Delta x^3)\\
%     \Leftrightarrow \dv{f(x)}{x}= \frac{f(x+\Delta x)-f(x-\Delta x)}{2\Delta x}+\mathcal{O}(\Delta x^3)\,,
% \end{gather}
% and by adding equations \eqref{eq:Taylor} we get the discrete and symmetric second-order derivative of $f(x)$
% \begin{gather}
%     f(x+\Delta x)+f(x-\Delta x) = 2f(x)+2\dv[2]{f(x)}{x}\Delta x^2+\mathcal{O}(\Delta x^4)\\
%     \Leftrightarrow \dv[2]{f(x)}{x} = \frac{f(x+\Delta x)-2f(x)+f(x-\Delta x)}{2\Delta x^2}+\mathcal{O}(\Delta x^4)\,.
% \end{gather}
% When using a uniform grid cell $x = \{x_i\,|\,i\in 1,\dots, \hat{N}\}$ where $\Delta x$ is the spacing, we can introduce the shorthand notation $f(x_i)\equiv f_i$ so that the above expressions further simplify to
\begin{gather}
    \dv{f_i}{x} = \frac{f_{i+1}-f_{i-1}}{2\Delta x}\label{eq:1stDeri_Discrete},\\
    \dv[2]{f_i}{x} = \frac{f_{i+1}-2f_i+f_{i-1}}{\Delta x^2}\label{eq:2ndDeri_Discrete}\,.
\end{gather}
After applying this discretization scheme to the energy domain of the NBDE with constant transport coefficients from equation (\ref{bose}), we get a system of $\hat{N}$ coupled differential equations:
\begin{equation}\label{eq:NBDE_Const_Discretized}
    \pdv{n_i}{t} = -v\frac{n_{i+1}-n_{i-1}}{2\Delta \epsilon}\,(1+2n_i)+D\frac{n_{i+1}-2n_i+n_{i-1}}{\Delta\epsilon^2}\,.
\end{equation}\\
This system of differential equations can now be integrated in time using the \textit{DifferentialEquations.jl} package \cite{rackauckas2017differentialequations} of the scientific programming language \textsc{Julia} \cite{Julia-2017}. However, since the time evolution of one individual cell $n_i$ for $i\in 1\dots \hat{N}$ depends on the values of its adjacent cells $n_{i\pm1}$ we have to incorporate the boundary conditions and introduce two additional cells with constant values as follows:
\begin{gather}
    \lim\limits_{\epsilon\rightarrow 0} n\,(\epsilon,t) = \infty \quad\rightarrow\quad n_{i=0}(t) = n_0(t) = 10^{10}\,,\\
    \lim\limits_{\epsilon\rightarrow \infty} n\,(\epsilon,t) = 0 \quad\rightarrow\quad n_{i=\hat{N}+1}(t) = n_{\hat{N}+1}(t) = 0.
\end{gather}
The value $\hat{N}$ limits the number of grid cells/the number of coupled differential equations that the program has to solve, and the divergent character of the solution on the left boundary is ensured by a large finite value. However, since both boundary conditions are only approximations, these will inevitably lead to erroneous behavior on and around the boundaries of the solution.

Furthermore, the implementation of the initial condition $n_\mathrm{i}(\epsilon) = n_\mathrm{i}^0\theta(1-\epsilon/Q_\mathrm{s})$ introduces an error at the cutoff $Q_\mathrm{s}$. This is because the exact analytical initial condition has a sharp discontinuity for $\epsilon = Q_\mathrm{s}$ but for the numerical solution, the cutoff needs to have a finite width, which is determined by the spatial resolution $\Delta \epsilon$.
% which can be seen in \ref{fig:InitialConditionSmoothed}.
The numerical initial condition is chosen to be symmetrically smoothed around the saturation momentum $Q_\mathrm{s}$ such that
\begin{gather}
    n_\mathrm{num,\,initial}(Q_\mathrm{s}-\Delta \epsilon) = n_\mathrm{i}^0,\\
        n_\mathrm{num,\,initial}(Q_\mathrm{s}) = n_\mathrm{i}^0/2,\\
    n_\mathrm{num,\,initial}(Q_\mathrm{s}+\Delta \epsilon) = 0.
\end{gather}
Even though we could further decrease the width of the slope by using an asymmetric approach and defining the value of $n_\mathrm{num,\,initial}$ at the saturation momentum as either $n_\mathrm{i}^0$ or $0$, the symmetric approach has proven more useful in minimizing the error around the cut-off during thermalization.
% (see \ref{Section:ErrorAnalysis}).
The actual numerical solution of the discretized NBDE for constant coefficients, equation (\ref{eq:NBDE_Const_Discretized}), will
be shown later, and compared to the exact analytical result.

We calculate the numerical particle distribution function $n_\mathrm{num}(\epsilon, t)$ for the same parameters as in our previous analytical solution shown in Fig.\,1 of \cite{mgw24}, with an equilibrium temperature that is roughly characteristic for the central temperature reached in a Pb-Pb collision at LHC energies following local thermalization. 
%Since microscopic calculations based on quantum chromodynamics are not available for the transport coefficients,
The values of the transport coefficients are determined as already outlined in \cite{mgw24} from the respective relations to the equilibrium temperature $T$, and the local bosonic equilibration time $\tau_\mathrm{eq}=4D/(9v^2)$ derived in \cite{gw18} at the gluon saturation scale $Q_\mathrm{s}$  as
\begin{equation}
 D=4\,T^2/(9\tau_\mathrm{eq})\,, \hspace{0.6cm} v=-4\,T/(9\tau_\mathrm{eq})\,.
 \label{Dv}
 \end{equation}
To obtain the relation between the transport coefficients and the local equilibration time, we have performed an asymptotic expansion of the error functions in the analytical solutions \cite{gw18} at the UV boundary $Q_\mathrm{s}$, whereas the fluctuation--dissipation relation $T=-D/v$ is a consequence of the equality of the stationary solution with the Bose--Einstein distribution as described before. 

For conserved energy density in the initial and the thermalized gluon distribution, the equilibrium temperature $T$ and the initial occupation were found to be related 
in \cite{jpb13} as $T=[15n_\text{i}^0/(4\pi^4)]^{1/4}Q_\mathrm{s}$, yielding $T\simeq443$\,MeV
for $Q_\mathrm{s}=1$\,GeV and $n_\text{i}^0= 1$, which we choose for convenience. This refers to an overoccupied system, as is typical for relativistic heavy-ion collisions at LHC energies. For example, the initial central temperature in $\sqrt{s_\text{NN}}=2.76$ TeV Pb-Pb collisions was determined to be $T_\text{ini}\simeq 480$ MeV \cite{hnw17} from bottomonium suppression in the quark-gluon plasma, corresponding to $n_\text{i}^0\simeq1.38$. This is significantly above the critical initial occupation $n_\text{i}^0=0.154$ where the particle-number density in equilibrium becomes equal to the one of a $\theta$-function initial distribution, $N_\text{eq}^{\mu=0}/N_\text{i}=6\zeta(3)(15/4)^{3/4}/[(n_\text{i}^0)^{1/4}\pi^3]=1$, thus defining the boundary from an under- to an overoccupied situation \cite{jpb12,blmt17}.

In principle, the transport coefficients  $v$ and $D$ that appear in the NBDE should be obtained self-consistently from the particle
distribution itself, as has been done in the numerical solution of a related transport equation \cite{blmt17}. However, their solution for the distribution functions that is valid
in the infrared up to $\epsilon\le 0.4\,Q_\text{s}$ compares well with our analytical NBDE result for constant coefficients. Hence, it appears reasonable to 
explore the consequences of an energy-dependence of the transport coefficients on the results without explicit consideration of the self-consistency.

For gluons, an experimental determination of the equilibration time is not possible. Instead, one has to rely on the comparison between model calculations and data. From coupled %kinetic equations \cite{fmr18} an upper limit of $\tau_\mathrm{eq}\lesssim 0.25$\,fm has been obtained, to be compared with an estimated freezeout time in central Pb-Pb collisions %at 
kinetic equations \cite{fmr18} an upper limit of $\tau_\mathrm{eq}\leq 0.25$\,fm has been obtained -- significantly shorter than the estimated freezeout time in central Pb-Pb collisions at LHC energies of $8-10$\,fm. However, these authors use a linear relaxation-time approximation in their numerical calculations of the collision term, such that a smaller value should be used in a nonlinear approach. Here we take $\tau_\text{eq}=0.13$\,fm, and obtain the transport coefficients from Eqs.\,({\ref{Dv}) as $D=0.67$\,GeV$^2$/fm and $v=-1.51$\,GeV/fm, which we use in the subsequent model calculations.

Whereas the initial condition in the exact calculation is a $\theta$-function up to the gluon saturation momentum $Q_\text{s}$, for the numerical solution we take a symmetrically smoothed box distribution as defined above with $n_\mathrm{i}^0 = 1$ and $Q_\mathrm{s} = \SI{1}{\GeV}$. For the numerical simulation, we use a grid with $\hat{N}=5000$ cells and a spatial resolution of $\Delta\epsilon = \SI{1E-3}{\GeV}$.

Plots of analytical \cite{mgw24} versus numerical results for constant coefficients look identical, corroborating the numerical solution method. A closer inspection of tiny differences between analytical and numerical approaches for constant coefficients confirms the actual validity and applicability of our numerical solution method. The small deviations are most pronounced at or around the discontinuities from the respective initial and boundary conditions at $\epsilon=Q_\text{s}=1$ GeV and the infrared boundary, where the infinite value has been replaced in the numerical solution by $10^{10}$. {However, these local deviations were found to decay in time and no oscillatory behavior or blow-ups of the solution are observed. The numerical particle distribution function $n(\epsilon, t)$ remains bounded and consistent through grid refinements, which indicates an empirically stable discretization method. }
We have evaluated the time evolution of the absolute error of the numerical solutions. At $Q_\text{s}$, the time-averaged error indicates that our numerical solution is accurate to $4-5$ decimal places when compared to the exact one. It is therefore likely justified to extend the solution method to the NBDE with energy-dependent coefficients.
{In the next section, a grid refinement of the numerical solution is also performed for energy-dependent transport coefficients, demonstrating the converging behavior of the numerical solution for $\Delta\epsilon\rightarrow 0$.}

\section{Energy-dependent transport coefficients}
\label{energy_dep}
In the physically more realistic case of energy-dependent transport coefficients, one has to expand the previous discretization of the NBDE for constant coefficients.
We shall neglect a possible time dependence of the transport functions and consider only their energy dependence. Using the full NBDE (\ref{nbde}) and the discretization scheme from Eqs.\, (\ref{eq:1stDeri_Discrete}) and (\ref{eq:2ndDeri_Discrete}), we are left with the discretized version of the nonlinear boson diffusion equation
\begin{equation}
\label{eq:NBDE_Full_Discrete}
\begin{split}
    \dv{n_i}{t} =& -\frac{v_{i+1}n_{i+1}-v_{i-1}n_{i-1}}{2\Delta \epsilon} (1+n_i)-v_in_i\frac{n_{i+1}-n_{i-1}}{2\Delta\epsilon}\\
    &+\frac{D_{i+1}-D_{i-1}}{2\Delta\epsilon}\frac{n_{i+1}-n_{i-1}}{2\Delta\epsilon}+D_i \frac{n_{i+1}-2n_i+n_{i-1}}{\Delta\epsilon^2}\,.
\end{split}
\end{equation}
Here, the distributions of the transport coefficients in energy space are discretized, as well as their derivatives.

We shall first discuss the energy dependence of the transport coefficients, followed by a discussion of the time-dependent thermalization process using energy-dependent coefficients as compared to the one for constant coefficients. Since no observables exist that allow for a direct measurement of gluon thermalization, there is no data available that could be used to infer the energy dependence of the transport coefficients. 
%On the theoretical side, their calculation from quantum chromodynamics is presently not feasible. 
However, we have used an ansatz for the energy dependence in the case of ultracold atoms \cite{lgw24}, where time-dependent thermalization is actually measurable. For gluons, we propose to proceed analogously: Since transport processes in the ground state should be minimized, the drift and diffusion terms are set to zero at the infrared boundary $\epsilon=0$, as well as for very large values, $\epsilon\rightarrow\infty$. We use an ansatz that is similar to the one in \cite{lgw24}
\begin{equation}\label{eq:EnergyDependence_General}
    f(\epsilon) = a\epsilon\,e^{-\epsilon/\epsilon_\mathrm{peak}},
\end{equation}
with positive constants $a>0$ and $\epsilon_\mathrm{peak}>0$. The normalization of the distribution is determined by $a$, while $\epsilon_\mathrm{peak}$ is the peak position in energy space. We normalize the functions representing both transport coefficients such that their mean values $\langle D\rangle$ and $\langle v\rangle$ are equal to $D, v$ from the previous constant-coefficient calculation. With the mean value for a function on a finite domain
\begin{equation}
    \langle f \rangle = \frac{1}{L} \int\limits_0^L f(x)\dd{x}\,,
\end{equation}
and inserting equation (\ref{eq:EnergyDependence_General}), we get the relation for the normalization constant
\begin{equation}
\begin{gathered}
    \langle f\rangle = \frac{1}{L}\int\limits_0^L a\epsilon\,e^{-\epsilon/\epsilon_\mathrm{peak}}\dd{\epsilon} =\frac{a \epsilon_\mathrm{peak}}{L} \left(\epsilon_\mathrm{peak}-e^{-L/\epsilon_\mathrm{peak}}(L+\epsilon_\mathrm{peak})\right)\\
    \Leftrightarrow a  = \underbrace{\frac{L}{\epsilon_\mathrm{peak}\left(\epsilon_\mathrm{peak}-e^{-L/\epsilon_\mathrm{peak}}(L+\epsilon_\mathrm{peak})\right)}}_{\equiv \tilde{a}}\langle f \rangle\,.
\end{gathered}
\end{equation}
Hence, the distributions of drift and diffusion in energy space simplify to
\begin{gather}
    D(\epsilon) = \hat{a}\langle D \rangle\epsilon_\mathrm{peak}e^{-\epsilon/\epsilon_\mathrm{peak}}\label{eq:D_EnergyDep}\\
    v(\epsilon) = \hat{a}\langle v \rangle\epsilon_\mathrm{peak}e^{-\epsilon/\epsilon_\mathrm{peak}}\label{eq:v_EnergyDep}\,,
\end{gather}
where the mean $\langle\cdot\rangle$ of the distributions is equal to the constant values from last sections $D = \SI{0.67}{\GeV^2/\fm}$ and $v = \SI{-1.51}{\GeV/\fm}$ with
\begin{equation}\label{eq:TransportCoeffRelation}
    \frac{D(\epsilon)}{v(\epsilon)} = \frac{\langle D\rangle}{\langle v\rangle} = -T\,.
\end{equation}
For a peak at $\epsilon_\mathrm{peak} = \SI{1.7}{\GeV}$, we display the distributions of $D(\epsilon)$ and $v(\epsilon)$ together with the equilibrium temperature $T$ in figure \ref{fig1}. The temperature is seen to be constant for every point in the energy domain.
\begin{figure}
	%\begin{widetext}
	\begin{center}
	\includegraphics[scale=0.54]{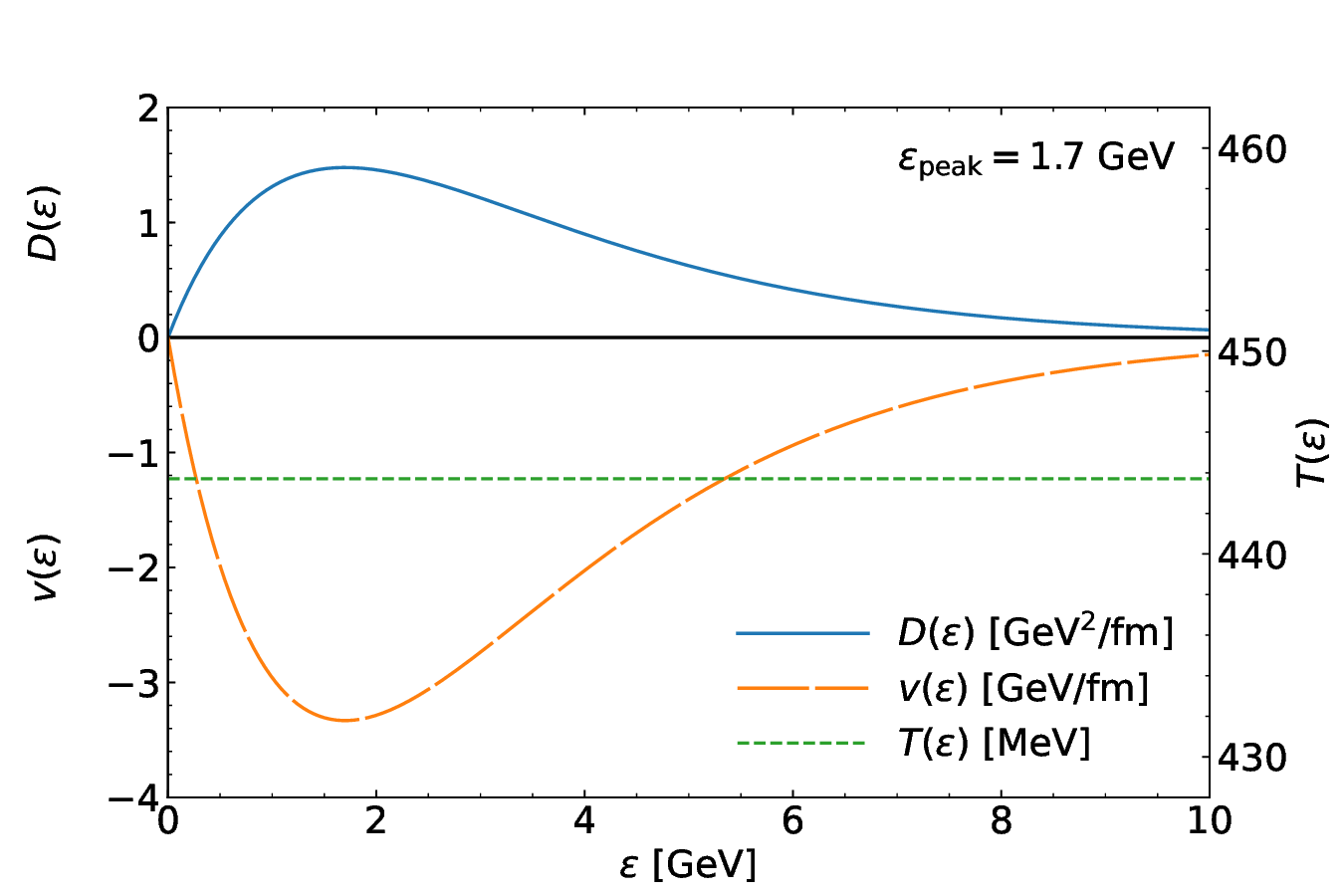}% Here is how to import EPS art
	\caption{\label{fig1}
Energy-dependent transport coefficients for an equilibrating system of massless gluons. 
%calculated from a Boltzmann-distribution ansatz. 
The upper solid curve is the diffusion coefficient $ D(\epsilon)$, the lower long-dashed curve is the drift term $ v(\epsilon)$, for an initial distribution $\theta(1-p/Q_\text{s})$ with  $p=|{p}|\equiv\epsilon$ for massless gluons, the gluon saturation momentum $Q_\text{s}\simeq 1$ GeV, and an equilibrium temperature $T=-D(\epsilon)/v(\epsilon)=443$ MeV that is constant across the whole energy domain (short-dashed line, right scale).}
	%\end{widetext}
	\end{center}
\end{figure}

We proceed to calculate the gluon occupation-number distribution functions with the above energy-dependent transport coefficients using the method-of-lines discretization from Eq.\,(\ref{eq:NBDE_Full_Discrete}) and solving the system of differential equations with \textsc{Julia} \cite{Julia-2017, rackauckas2017differentialequations}. The results are compared  for two values of time -- 0.1 fm and 0.3 fm -- with the previous ones for constant transport coefficients in Fig.\,\ref{fig2},
where the time-dependent thermalization for both energy-dependent and constant coefficients is shown.
\begin{figure}
	%\begin{widetext}
	\centering
	\includegraphics[scale=0.6]{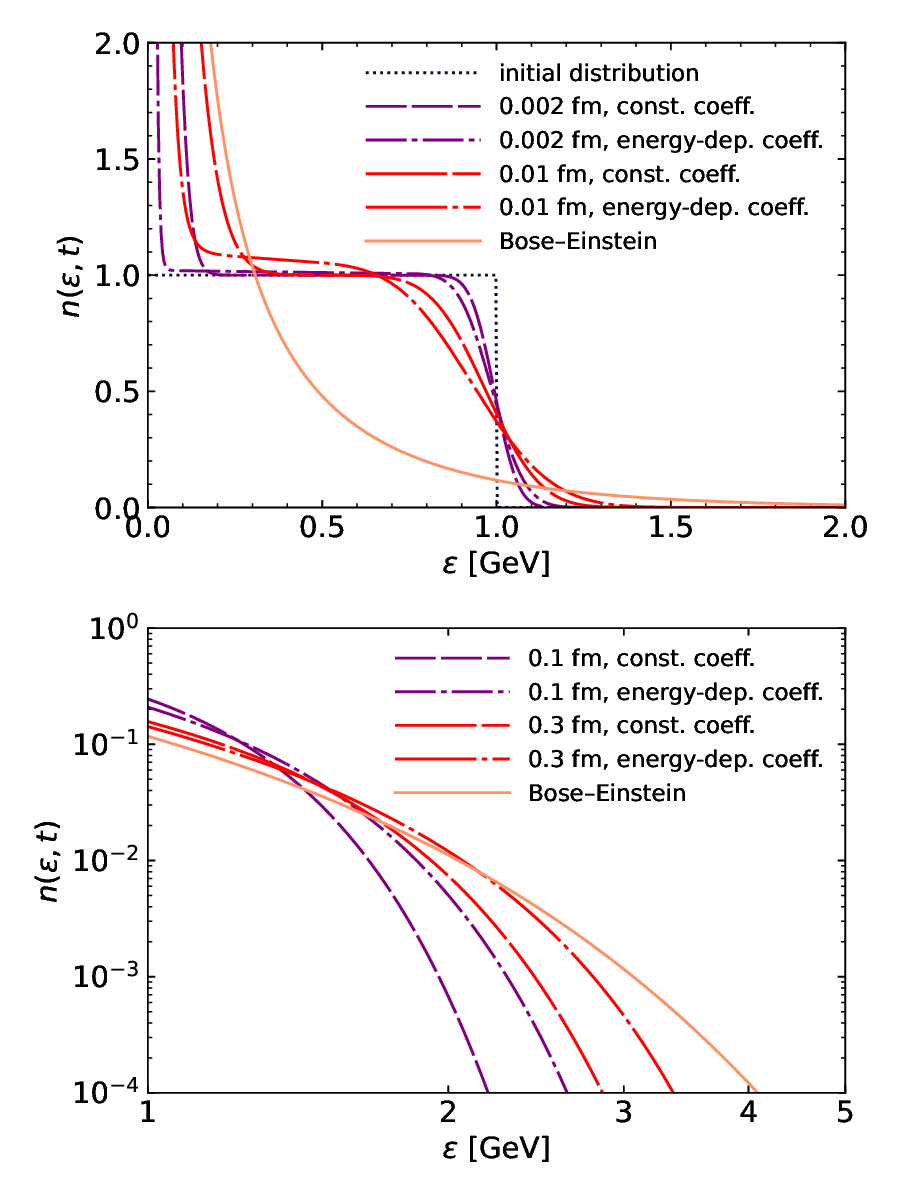}% Here is how to import EPS art
	\caption{
Comparison of NBDE-solutions for constant (short-dashed at short times, long-dashed at larger times) and energy-dependent transport coefficients (dot-dashed). The energy dependence has the exponential form shown in Fig.\,\ref{fig1} with its peak at $\epsilon_\mathrm{peak} ={1.7}$ {GeV} and is chosen such that the mean value of the energy-dependent transport coefficients equals the value for constant coefficients: $\langle D(\epsilon) \rangle = D = {0.67}$ GeV$^2$/fm and $\langle v(\epsilon) \rangle = v = {-1.51}$ GeV/fm. In the top frame, the thermalization in the infrared region is seen to be faster for constant transport coefficients. At the saturation momentum $Q_\mathrm{s}$, both solutions thermalize almost equally fast, while for the ultraviolet region, the energy-dependent transport coefficients provide a much faster thermalization than the constant ones. This is clearly visible in the double logarithmic plot, bottom frame.} 
\label{fig2}
\end{figure}
The transport coefficients are $\langle D \rangle = \SI{0.67}{\GeV^2/\fm}$ and $\langle v\rangle = \SI{-1.51}{\GeV/\fm}$ and the energy distribution of the transport coefficients is as in Fig.\,\ref{fig1}, with a maximum at $\epsilon_\mathrm{peak} = \SI{1.7}{\GeV}$. The peak is placed in the ultraviolet energy range in order to achieve a faster (slower) thermalization in the ultraviolet (infrared) region compared to the constant coefficients. Later, it will be shown that -- depending on the peak position -- this effectively hinders the formation of an overshoot in the time-dependent condensate fraction.

The time evolution of the occupation-number distribution functions for both, constant and energy-dependent coefficients is
compared for nine timesteps in separate plots, Fig.\,3,
with the same legend in the four frames, and the same transport coefficients as in Figs.\,1, 2. Figs.\, 3\,(a), (c)  on the left show the analytical NBDE solutions for constant transport coefficients, while Figs.\, 3\,(b), (d) on the right show the numerical NBDE solutions for the energy-dependent transport coefficients.
\begin{figure}
	\centering
  \includegraphics[scale=0.66]{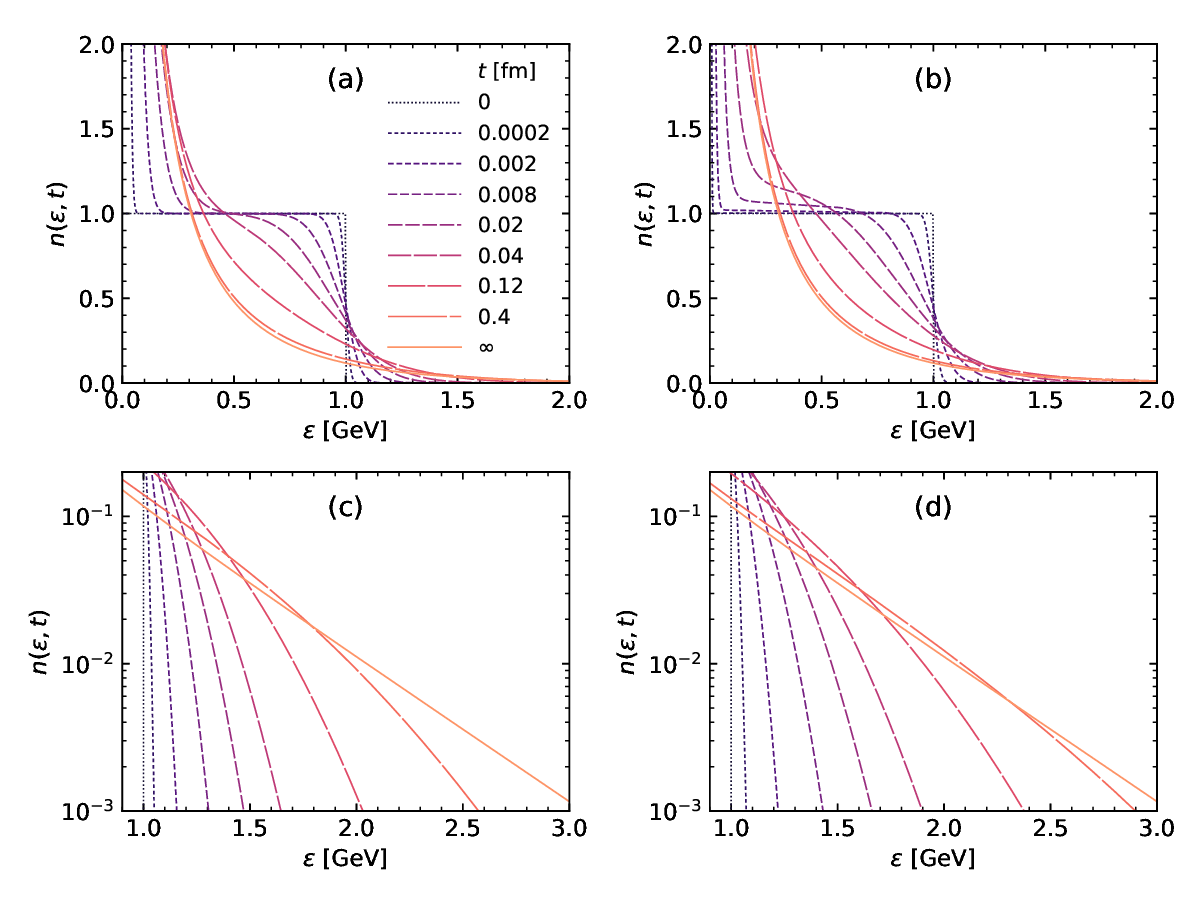}
\caption{Nonequilibrium evolution of a massless gluon system using the solutions of the nonlinear boson diffusion equation with constant transport coefficients (a), (c) are compared to the numerical solutions (b), (d) of the NBDE with energy-dependent transport coefficients. The initial state is $\theta(1-p/Q_\text{s})$ with the gluon saturation momentum $Q_\text{s}\simeq 1$ GeV. The constant transport coefficients are $D\equiv\langle D\rangle=0.67\,\text{GeV}^2/\text{fm}$ and $v\equiv\langle v\rangle=-1.51\,\text{GeV}/\text{fm}$.
The equilibrium temperature is $T=-D/v=443\,\text{MeV}$, with the corresponding Bose--Einstein distribution (solid curves). Time-dependent single-particle occupation-number distribution functions are shown at the timesteps shown in (a), with increasing dash lengths, same steps in all four frames. The calculations for constant coefficients in (a), (c) show identical results, but (c) zooms into UV energies; accordingly for the numerical results with energy-dependent coefficients shown in in (b), (d). }
\label{fig3}       % Give a unique label
\end{figure}

Apart from the different solution methods due to the energy dependence, all four subfigures use the same parameters: The initial distribution is a $\theta$-function for an overoccupied system with $n_\mathrm{i}^0 = 1$ and saturation momentum $Q_\mathrm{s}$. The equilibrium distribution is a Bose--Einstein distribution with temperature $T = -D/v = \SI{443}{\MeV}$. The parameters of the numerical calculation are $L = \SI{10}{\GeV}$ with a cell width of $\Delta\epsilon = \SI{2E-3}{\GeV}$ and the value $n(\epsilon = 0) = 10^{10}$ for the infrared boundary for $t>0$.

Comparing Figs.\,3\,(a) and (b), a much slower thermalization in the infrared region is observed for the energy-dependent coefficients: It takes nearly an order of magnitude longer to get the same degree of thermalization for the energy-dependent coefficients as for the constant ones due to the small values of $D(\epsilon)$ and $v(\epsilon)$ in the infrared region.

The thermalization for the intermediate energy region around the saturation momentum $Q_\mathrm{s}$ behaves very similar for constant and energy-dependent coefficients, and the edges of the initial distribution are continuously smeared out in both cases.
A close look at Figs.\,3\,(c) and (d) confirms, however, that the model with energy-dependent transport coefficients causes a faster thermalization in the ultraviolet region -- as has already been indicated for two timesteps in Fig. \,\ref{fig2}. This difference becomes more evident for later times, when the thermal tail of the distribution develops much faster for energy-dependent coefficients than for constant coefficients, and it will prevent an overshoot of the time-dependent condensate fraction above the equilibrium value.

{Since there exists no analytical solution of the NBDE with energy-dependent transport coefficients, an estimate of the (mean) local error of the numerical solution becomes more difficult. However, we can calculate an upper bound for the error by comparing the numerical solutions for different grid resolutions $\Delta\epsilon$. First, we start by computing the mean absolute difference of the numerical solutions for a fine grid $n^{(\Delta\epsilon)}(\epsilon, t)$ with spacing $\Delta\epsilon$ and a coarser grid $n^{(2\Delta\epsilon)}(\epsilon, t)$ with spacing $2\Delta\epsilon$
\begin{equation}
    \langle|n^{(\Delta\epsilon)}(\epsilon, t) -n^{(2\Delta\epsilon)}(\epsilon, t)|\rangle_\epsilon = \frac{1}{N}\sum\limits_{i = 1}^{N} |n^{(\Delta\epsilon)}(\epsilon_i, t)-n^{(2\Delta\epsilon)}(\epsilon_i, t)|,
\end{equation}
with the number of cells $N = L/2\Delta\epsilon$ of the coarse grid and $\epsilon_i = 2i\Delta\epsilon$.
By calculating an additional time average over $T$ different times, we are left with a total mean of the absolute difference $E(\Delta\epsilon)$ between the two solutions with different grid resolutions
\begin{equation}
    E(\Delta\epsilon) = \langle|n^{(\Delta\epsilon)}(\epsilon, t) -n^{(2\Delta\epsilon)}(\epsilon, t)|\rangle_{\epsilon,t} = \frac{1}{T}\sum\limits_{i = 1}^T\langle|n^{(\Delta\epsilon)}(\epsilon, t_i) -n^{(2\Delta\epsilon)}(\epsilon, t_i)|\rangle_\epsilon.
\end{equation}
We attribute the mean difference $E$ of the two solutions with grid sizes $\Delta\epsilon$ and $2\Delta\epsilon$ to the finer grid size $\Delta\epsilon$, and repeate this process for different $\Delta\epsilon$.
{The mean absolute difference gets smaller as $\Delta\epsilon\rightarrow 0$. An almost linear relationship between mean absolute difference $E(\Delta\epsilon)$ and grid size $\Delta\epsilon$ is found in a  double logarithmic plot, illustrating the empirical convergence of the numerical solution under grid refinements. Therefore, an estimate of the mean local error of the numerical solution $n(\epsilon, t)$ with energy-dependent transport coefficients can now be made for the finest grid size $\Delta\epsilon = \SI{2E-3}{\GeV}$ with $N = 5000$ cells, by taking the value of $E(\SI{2E-3}{\GeV})\approx \SI{6.4E-3}{}$ as a conservative upper bound for its mean local error.}

\section{Time-dependent condensate fraction}
\label{cond}
\subsection{Condensate fraction with constant transport coefficients}
Conservation of particle number for elastic collisions in overoccupied systems requires that the excess particles in the initial distribution will have to move into a condensed state during the time evolution.  The condensate fraction is the time-dependent number of particles in the condensed state $N_\text{c}(t)$ relative to the number of particles $N_\text{i}$ in the initial state. Here, $N_\text{c}(t)$ is calculated from the difference between the number of initial particles $N_\mathrm{i}$ and the integrated particle number $N(t)$ from the NBDE solution $n\,(\epsilon,t)$  at time $t$:
\begin{equation}\label{eq:CondensateFraction}
    \frac{N_\mathrm{c}(t)}{N_\mathrm{i}} = \frac{N_\mathrm{i}-N(t)}{N_\mathrm{i}} = 1-\frac{N(t)}{N_\mathrm{i}}\,.
\end{equation}
The number of particles at any given time is an integral over the particle distribution function $n\,(\epsilon,t)$ and the density of states for a relativistic system $g(\epsilon)\propto \epsilon^2$  \cite{fuku17} 
\begin{equation}
    N(t) = \int\limits_0^\infty g(\epsilon)n\,(\epsilon,t)\dd{\epsilon}\,.
    \label{nopart}
\end{equation}
There exists an upper limit for the condensate fraction which is given by the ratio of the number of particles in the thermal Bose--Einstein distribution at temperature $T$ and the initial particle number of gluons in the nonequilibrium box distribution. It is obtained as
\begin{equation}\label{eq:CondensateFractionLimit}
\lim\limits_{t\rightarrow\infty} \frac{N_\mathrm{c}(t)}{N_\mathrm{init}} = 1-\left(\int\limits_0^{\infty } \frac{\epsilon^2}{e^{-v \epsilon/D}-1}\dd{\epsilon}\right)\Bigg/\left(\int\limits_0^{Q_\mathrm{s}} \epsilon^2 n_{\mathrm{i}}^0\dd{\epsilon}\right) = 1+\frac{6 D^3 \zeta(3)}{v^3 n_\mathrm{i}^0 Q_\mathrm{s}^3}\,,
\end{equation}
where $\zeta(3)\approx 1.2$ is the Riemann $\zeta$-function. 
The condensate fraction based on the numerical NBDE-solution is calculated as for the analytical one using Eq.\,(\ref{eq:CondensateFraction}).
% However, now the integral over the particle distribution has to be discretized so that the particle number $N(t)$ is calculated as follows
% \begin{equation}\label{eq:CondensateFractionNumerical}
%     N(t) \propto \int\limits_0^\infty \epsilon^2 n\,(\epsilon,t)\dd{\epsilon}\approx \int\limits_0^{\hat{N}\Delta\epsilon} \epsilon^2 n\,(\epsilon,t)\dd{\epsilon} \approx \sum\limits_{i = 1}^{\hat{N}} \epsilon_i^2\,n_\mathrm{num}(\epsilon_i,t)\,.
% \end{equation}
% Here, the upper limit of the first integral is approximated with the finite right bound $\hat{N}\Delta\epsilon=L$ of the grid in the first step, while in the second step, the continuous integral is approximated by a discrete summation over all cells.
Calculating the condensate fraction for each time step, and with the same constant transport parameters $v, D$ as in Fig.\,\ref{fig2} yields the result for $N_\mathrm{c}(t)/N_\mathrm{i}$, see Fig. \ref{fig4}.
\begin{figure}
	%\begin{widetext}
	\centering
	\includegraphics[scale=0.6]{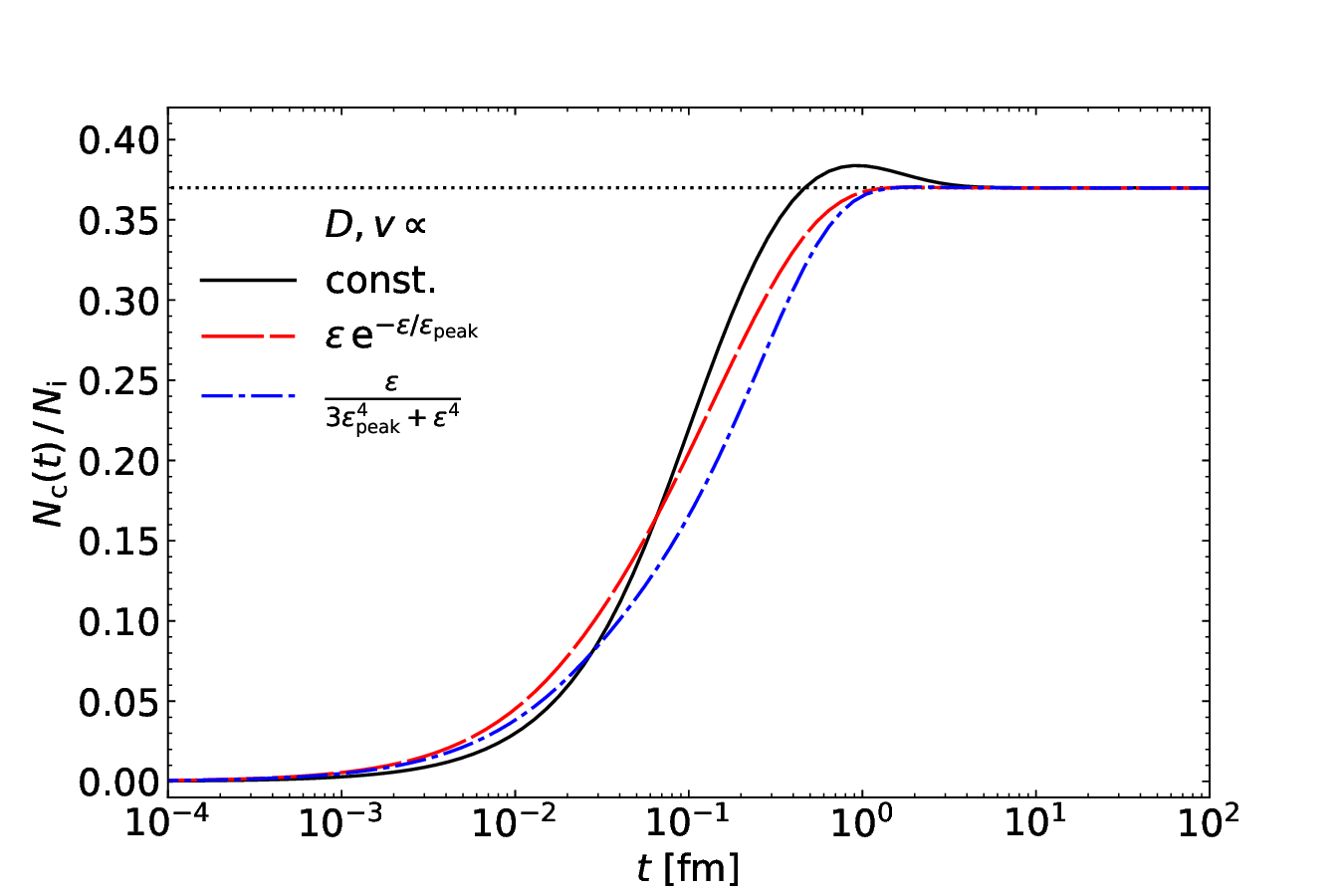}% Here is how to import EPS art
	\caption{\label{fig4}
Gluon-condensate fraction in an overoccupied system using constant coefficients (solid curve) compared with the result for two different energy-dependent transport coefficients with $\epsilon_\mathrm{peak} = 1.7$ GeV, an exponential distribution of the energy dependence of $D$ and $v$ as shown in Fig.\,\ref{fig1} (dashed curve) and a rational distribution (dot-dashed curve). The condensate fraction is calculated by imposing particle-number conservation and using Eqs.\,(\ref{eq:CondensateFraction}) and (\ref{nopart}). The mean values of the distributions of the energy-dependent coefficients are equal to $D = 0.67$ GeV$^2$/fm and $v = -1.51$ GeV/fm, as used for constant coefficients. The energy dependencies of $D$ and $v$ are adjusted to counteract the fast (slow) thermalization in the infrared (ultraviolet) region for constant transport coefficients and thus prevent an overshoot above the equilibrium limit $0.369958$ (dotted line) of the condensate fraction.}
\end{figure}
After some initialization time, the condensate fraction is seen to grow as expected, dashed curve. But instead of approaching the equilibrium limit given by Eq.\,(\ref{eq:CondensateFractionLimit}) from below, it slightly overshoots and starts to approach it from above. This is a direct result of the assumption of constant transport coefficients: The thermalization in the infrared region is too fast compared to the one in the ultraviolet region, and hence, too many particles are being pushed into the condensate, thus overshooting the thermal limit. Later, it will be shown that with properly adjusted energy-dependent transport coefficients, a faster (slower) thermalization in the ultraviolet (infrared) region will prevent this overshoot and cause condensate formation to approach its thermal limit from below, as is physically required.

Whereas the calculation of the condensate fraction for constant coefficients shown in Fig.\,\ref{fig4} is performed fully numerically using \textsc{Julia}, we had already calculated it using the exact analytical NBDE-solutions for $n\,(\epsilon,t)$ in \cite{mgw24} for the same transport parameters, seeFig.\,2 in that reference. There, a second -- almost invisible -- overshoot of the equilibrium  limit appears at $t\simeq 5$\,fm. This tiny bump is due to a so-called catastrophic cancellation that occurs when subtracting two closely neighbouring numbers that have rounding errors \cite{CancellationGoldberg, CancellationMuller2018}. In our case, this happens for differences of the error functions that appear in the exact analytical results for the time-dependent partition function and its energy-derivative given in \cite{mgw24}. The integral over the Gauss distribution that defines the error function
\begin{equation}
    \erf(x) = \frac{2}{\sqrt{\pi}}\int\limits_0^\infty e^{-x^2}\dd{x}
\end{equation}
cannot be computed analytically in a closed form, but rather has to be evaluated numerically. The result is then rounded to the nearest floating point number, causing an amplification of the rounding errors \cite{CancellationMuller2018} when computing the difference $\erf(x)-\erf(y)$ for $x,y\gg 1$ since $\erf(x) \rightarrow 1$, so that the accurate bits of the error functions cancel each other and what remains are the rounding errors with which they were initially computed \cite{CancellationGoldberg}.

These errors are therefore merely of numerical nature when computing the values of the error functions and do not compromise the validity of the exact analytical solution of the NBDE. They only appear for very large arguments of the error functions, $x\gg1$, corresponding to $\epsilon\gg 1$ and $t\gg 1$ fm in the NBDE context. Hence, a direct comparison between the analytical	 and numerical solution is still justified and viable for values of $t$ and $\epsilon$ in the physically interesting and relevant regions.
Since the fully numerical calculation of the condensate fraction does not make use of error functions, the second bump after the initial overshoot is no longer present.

\subsection{Condensate fraction with energy-dependent transport coefficients}

The condensate fraction for the NBDE with energy-dependent coefficients is calculated in a similar fashion as for the case of constant coefficients. By imposing particle number conservation for elastic collisions, we attribute the difference of particle numbers at time $t$ and time $t= 0$ to the formation of a condensate and calculate the condensate fraction using Eq.\,(\ref{eq:CondensateFraction}).

{In figure \ref{fig4}, the condensate fractions of two different energy dependencies of the transport coefficients (dashed and dash-dotted curve) are compared with the condensate fraction for constant coefficients (solid curve). The dashed curve uses the exponential energy dependence of the transport coefficients shown in Fig\, \ref{fig1}, whereas the thermalization of its respective particle distribution function $n\,(\epsilon,t)$ has already been shown with the one for constant coefficients in Fig\,\ref{fig3}.
The dash-dotted curve displays the condensate fraction for a rational energy dependence of the transport coefficients
\begin{equation}
    D(\epsilon) =  \frac{a\epsilon}{3\epsilon_\mathrm{peak}^4+\epsilon^4},
\end{equation}
where the distribution for $v(\epsilon)$ is chosen accordingly such that Eq.\,(\ref{eq:TransportCoeffRelation}) holds, $a$ is a normalization constant and the denominator ensures the position of the maximum of the distribution at $\epsilon = \epsilon_\mathrm{peak}$.}

For both the exponential and rational distributions of drift and diffusion, the mean values are equal to the constant values $D = \SI{0.67}{\GeV^2/\fm}$ and $v = \SI{-1.51}{\GeV/\fm}$. The initial distribution is a box distribution with $n_\mathrm{i}^0 = 1$ so that the equilibrium value of the condensate fraction is calculated again with Eq.\,(\ref{eq:CondensateFractionLimit}) as $0.369958$.

In order to avoid the overshoot of the condensate fraction that occurs for constant coefficients,
we place the peak of the energy distributions $\epsilon_\mathrm{peak}$ of the transport coefficients in the ultraviolet region, so that the thermalization for high (low) energies is sped up (slowed down). With this adjustment, the condensate fraction for energy-dependent transport coefficients approaches the equilibrium value from below, as required.
Due to the placement of the peak in the ultraviolet, the overshoot of both energy dependencies is no longer observed, whereas their only difference lies in the behavior of the curve before it approaches the equilibrium value, which is a direct effect of the two different shapes of the energy dependencies.

In Fig\,\ref{fig5}, the influence of different peak locations $\epsilon_\mathrm{peak}$ in the energy dependence of the transport coefficients is shown, here for the exponential function as displayed in Fig.\,\ref{fig1}.
\begin{figure}
	%\begin{widetext}
	\centering
	\includegraphics[scale=0.6]{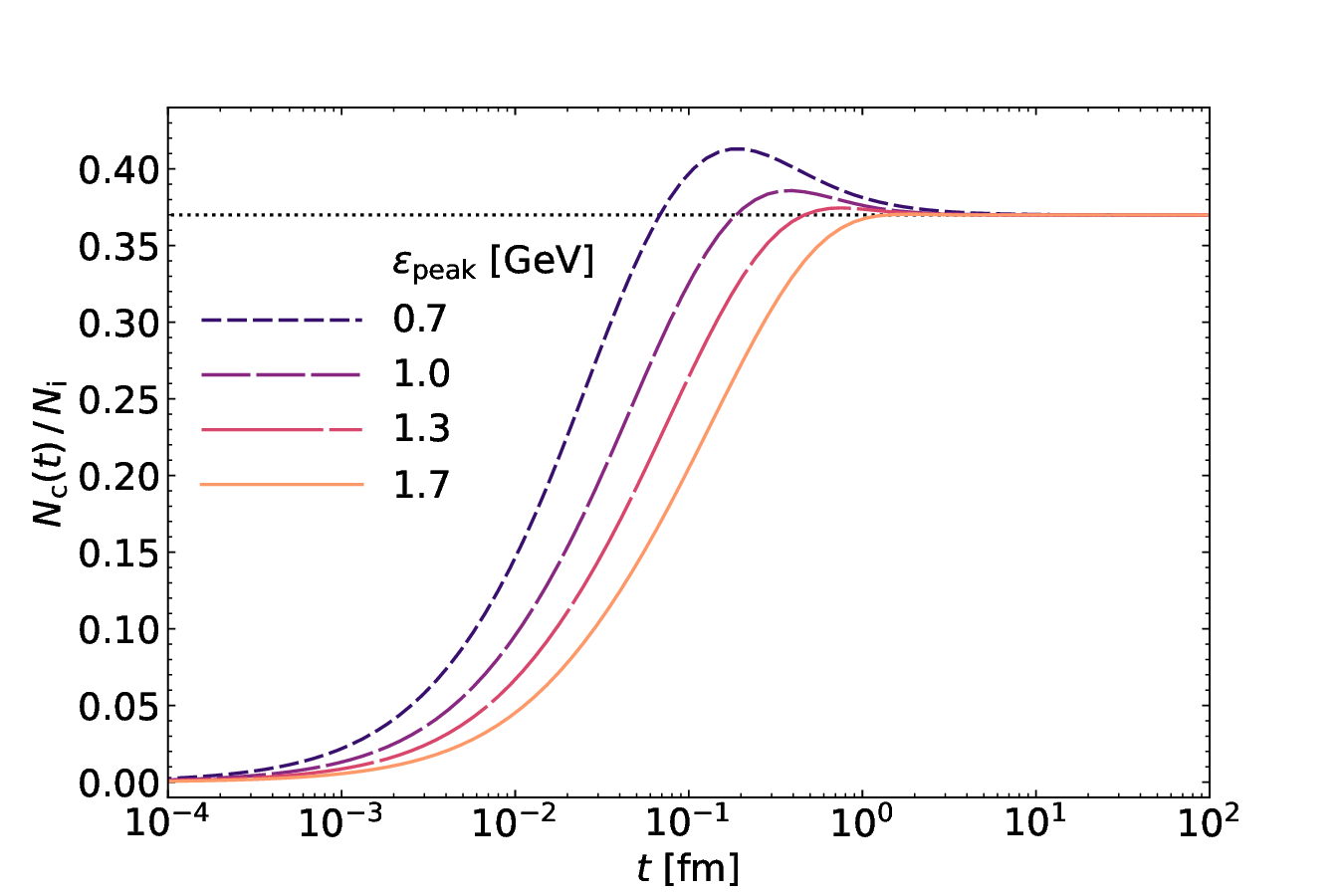}% Here is how to import EPS art
	\caption{\label{fig5}
{Effect of different peak locations $\epsilon_\mathrm{peak}$ in the energy distribution of the transport coefficients (Fig\,\ref{fig1}) on the condensate fraction. The further the maximum lies in the ultraviolet region $\epsilon_\mathrm{peak}>Q_\mathrm{s} = \SI{1}{\GeV}$, the smaller the overshoot gets: The thermalization in the infrared region is slowed down and the thermal tail in the ultraviolet develops faster, causing a balanced thermalization across the whole energy domain and
preventing an overshoot. The mean values for the transport coefficients are in all cases $\langle D\rangle = 0.67$ GeV$^2$/fm and $\langle v\rangle = -1.51$ GeV/fm. }
	}
\end{figure}
The mean values of the energy-dependent transport coefficients are equal in all cases, only the peak positions $\epsilon_\mathrm{peak}$ have been modified, such that the degree of the overshoot above the equilibrium value changes: The further the peak lies in the ultraviolet region, the smaller the overshoot becomes until it eventually vanishes completely. This is as expected since, by placing the peak in the ultraviolet, the thermalization for high-energy modes is accelerated, whereas the thermalization in the infrared is slowed down, thus preventing an overshoot. On the other hand, by placing the peak at energies smaller than the saturation momentum $\epsilon_\mathrm{peak}<Q_\mathrm{s}$ and thus amplifying the thermalization rate in the infrared, the overshoot can be enhanced and becomes even larger than the overshoot for constant transport coefficients.

Another effect that becomes visible when comparing the condensate fraction for different $\epsilon_\mathrm{peak}$ is that the location of the peak also influences the initialization time $\tau_\mathrm{ini}$ -- the time it takes for the condensation to begin -- and the equilibration time $\tau_\mathrm{eq}$ of the condensate. A peak at lower energies $\epsilon_\mathrm{peak}<Q_\mathrm{s}$ causes a faster thermalization, such that condensation starts earlier.

\section{Conclusions}
\label{conc}
We have studied the thermalization of gluons in the initial stages of relativistic heavy-ion collisions through elastic gluon scatterings  analytically and numerically by using a nonlinear partial differential equation for the single-particle occupation-number distribution $n\,(\epsilon,t)$ that properly accounts for the bosonic enhancement in the infrared domain.
This nonlinear boson diffusion equation had been derived earlier from the quantum Boltzmann equation, and
solved analytically for $\theta$-function initial conditions in the case of constant diffusion and drift coefficients without \cite{gw18} and with boundary conditions at the singularity \cite{gw22,mgw24}.

In the present work, a numerical method for solving the NBDE for arbitrary initial conditions has been presented, and its reliability is confirmed through a comparison with the exact analytical solution. The errors of the numerical solution are shown to be of the order of $10^{-4}\text{--} 10^{-6}$ with a time-averaged error at the gluon saturation momentum of $\SI{7.2E-6}{}$. Additionally, since the numerical solution does not make use of error functions, no cancellation error for large $t$ and $\epsilon$ occurs.

The numerical solution has been extended to accommodate the energy dependence of the transport coefficients that is present in the full NBDE. An energy dependence of $D$ and $v$ makes the model physically more realistic since transport processes should be minimized in the ground state, as well as for $\epsilon\rightarrow\infty$.
The energy dependencies of the transport coefficients have been modeled as exponential functions, as previously proposed for the case of cold atoms \cite{lgw24}, and their peaks were set to be in the ultraviolet region. As a consequence, the thermalization for the low-energy modes is slowed down, while the thermalization for the high-energy modes is accelerated.
The time evolution of the occupation-number distributions was then compared for the constant and energy-dependent transport coefficients: The solution for constant coefficients thermalizes much faster in the infrared region, whereas the solution for energy-dependent coefficients causes a more rapid thermalization in the ultraviolet region.

As in our previous work on exact solutions of the NBDE, we have calculated the time-dependent gluon-condensate formation through elastic scatterings from the condition of particle-number conservation, requiring that excess particles in overoccupied systems move into the condensed state. However, we now use the numerical solution scheme based on the programming language \textsc{Julia} that allows us to compute not only the solutions for constant, but also for energy-dependent transport coefficients.

In the case of constant transport coefficients, the time-dependent condensate fraction approaches the equilibrium limit, but slightly overshoots it before reaching it from above: Thermalization in the infrared energy region is too fast compared to the one in the ultraviolet domain. Only when the UV-tail of the distribution function starts to approach the Bose--Einstein tail from below, the number of free particles increases and the number of particles in the condensate decreases. Thus, the condensate fraction decreases and its equilibrium limit is approached from above. The equilibrium limit is reached for $t\rightarrow \infty$ when the ultraviolet energy modes have thermalized and the time-dependent distribution becomes equal to the Bose--Einstein distribution.
The numerical result for constant coefficients agrees with our earlier result that was based on the exact analytical calculation
of the distribution function, except for a minor numerical deviation due to the evaluation of the error functions in the exact result.

The condensate fraction of the NBDE with energy-dependent transport coefficients does not show an overshoot, but instead the equilibrium limit is approached from below as is physically required: The faster (slower) equilibration in the ultraviolet (infrared) region results in a more balanced thermalization over the whole energy domain and a continuously decreasing number of free particles in the thermal cloud with the proper parabolic density of states for gluons in relativistic heavy-ion collisions. Due to particle-number conservation for elastic collisions, the decreasing number of free particles results in an increasing number of particles in the condensate, so that eventually the equilibrium limit is approached from below without any initial overshoot.

By adjusting the peak  of the energy distribution, we have modified the overshoot of the condensate fraction, the initialization time of the condensate $\tau_\mathrm{ini}$ and the equilibration time $\tau_\mathrm{eq}$. With $\epsilon_\mathrm{peak}$ in the infrared region, the thermalization of the low-energy (high-energy) modes is even faster (slower) than for the constant coefficients, causing an increased thermalization of the initial particles that occupy the low-energy region with $\epsilon<Q_\mathrm{s} = \SI{1}{\GeV}$ and therefore smaller values of $\tau_\mathrm{eq}$ and $\tau_\mathrm{ini}$.  Our extension of the numerical solution to incorporate energy-dependent transport coefficients thus allows for a physically more realistic description of the thermalization of gluons. For a modified energy-dependent parametrization of the transport coefficients with same peak position $\epsilon_{peak}$, the overshoot in the condensate fraction also disappears, although a slightly different approach to the equilibrium value of the condensate fraction is observed.

In this work, we have focussed on elastic gluon collisions with vanishing chemical potential and the associated formation of a transient gluon condensate, whereas gluon thermalization through inelastic collisions had been considered earlier, albeit for constant transport coefficients only \cite{gw22a}. Even though a gluon condensate was initially an intriguing possibility \cite{jpb12, jpb13}, it has been shown that inelastic radiative processes occur on a much smaller time scale than elastic ones. The fast radiation of soft gluons causes the excess particles in overoccupied systems to be quickly suppressed and therefore, hinders the formation of a transient gluon condensate \cite{blmt17}.

Nevertheless, it is useful and interesting to explore the mechanisms that drive the thermalization of bosons and of time-dependent Bose--Einstein condensation. In particular, the nonlinear boson diffusion equation and its exact and numerical solutions for constant and energy-dependent transport coefficients, respectively, provide a framework that has been used successfully to account for actual data on the time-dependent condensation of ultracold atoms such as $\ce{^{23}Na}$ \cite{sgw21} and $\ce{^{39}K}$ \cite{kgw22,lgw24}, albeit in a much lower energy regime, and with initial conditions that are significantly different from the present approach to local gluon thermalization in relativistic heavy-ion collisions.
%\newpage
%\bigbreak
%\noindent
%\newpage
%\noindent
%{Acknowledgements}\\
%...\\

\bibliographystyle{elsarticle-num}

\end{document}